\documentclass[twocolumn,amsmath,amssymb,aps,pra,floatfix]{revtex4}

\usepackage{graphicx}
\usepackage{dcolumn}
\usepackage{bm}
\usepackage{hyperref}
\usepackage{xcolor}
\usepackage[utf8]{inputenc}
\usepackage{amsmath,bm}
\hypersetup{
    colorlinks,
    allcolors={blue!80!black},
    linktoc=all
}
\usepackage[sort&compress]{natbib}
\newcommand{\mathleft}{\@fleqntrue\@mathmargin\parindent}

\begin{document}


\title{Quantum noise spectra for periodically-driven cavity optomechanics}

\author{E. B. Aranas}
 \email{erika.aranas.14@ucl.ac.uk}
\author{M. Javed Akram}
\affiliation{
Department of Physics and Astronomy, University College London, Gower Street, London WC1E 6BT, United Kingdom
}
\author{Daniel Malz}
\affiliation{
Cavendish Laboratory, University of Cambridge, Cambridge CB3 0HE, United Kingdom \\}
\author{T. S. Monteiro}
\affiliation{
Department of Physics and Astronomy, UCL, Gower Street, London WC1E 6BT, United Kingdom\\
}

\date{\today}

\begin{abstract}

A growing number of experimental set-ups in cavity optomechanics exploit periodically driven fields. However, such set-ups are not amenable to analysis using simple, yet powerful, closed-form expressions of linearized optomechanics, which have provided so much of our present understanding of experimental optomechanics. In the present paper, we formulate a new method to calculate quantum noise spectra in modulated optomechanical systems, which we analyze, compare, and discuss with two other recently proposed solutions: we term these (i) frequency-shifted operators (ii) Floquet \cite{Malz2016.Nunn} and (iii) iterative analysis \cite{Aranas2016.Mont}. We prove that (i) and (ii) yield equivalent noise spectra, and find that (iii) is an analytical approximation to (i) for weak modulations. We calculate the noise spectra of a doubly-modulated system describing experiments of levitated particles in hybrid electro-optical traps. We show excellent agreement with Langevin stochastic simulations in the thermal regime and predict squeezing in the quantum regime. Finally, we reveal how experimentally inaccessible spectral components of a modulated system can be measured in heterodyne detection through an appropriate choice of modulation frequencies.

\end{abstract}

\maketitle


\section{\label{introduction}Introduction}

The last ten years have witnessed an impressive raft of experimental breakthroughs in the field of cavity quantum optomechanics \cite{Aspelmeyer2014}.  Despite the enormous diversity of experimental set-ups (including membranes, microtoroids, photonic crystal microcavities and levitated nanoparticles among others), most experiments are  amenable to analysis by means of the linearized theory of optomechanics. Through its well-established analysis in frequency space \cite{Aspelmeyer2014,Bowenbook} one may  obtain the  {\em quantum noise spectra}, in other words, the spectra of fluctuations (whether quantum or classical) of optical and mechanical modes subjected to thermal and optical noises from the environment. This enabled valuable insights on the physics underlying optomechanical cooling, strong coupling regimes, optical and mechanical squeezing, quantum back-action, as well as an understanding of the Standard Quantum Limit (SQL) of optomechanical displacement sensing \cite{Aspelmeyer2014,Bowenbook}. Hence, the analysis of quantum noise spectra from linearized optomechanical theory has become a ubiquitous tool of optomechanics.

Recently however, a number of experimental set-ups have involved periodically driven fields. Here we do not allude to classical feedback fields,  but rather to scenarios where cavity driving fields or other trapping fields are harmonically modulated in order to, for instance, generate mechanical squeezing \cite{Schwab,Sillanpaa,Teufel} or even to simply improve the trapping and cooling \cite{Millen2015,Fonseca2016.Bark,Aranas2016.Mont} of levitated optomechanical systems.  In such cases, even in regimes where nonlinearities are entirely absent from dynamics, one may no longer adapt the textbook closed-form mathematical expressions for quantum noise spectra.

An optomechanical system comprising a single optical cavity mode coupled to a mechanical oscillator is described by the well-known Hamiltonian \cite{Aspelmeyer2014,Bowenbook}:
\begin{equation}
{\hat{H}} = -\Delta{\hat a}^\dagger {\hat a} +
            \omega_{\textrm M} {\hat b}^\dagger {\hat b} +
            g ({\hat a}^\dagger+ {\hat a}) ({\hat b}^\dagger+ {\hat b}),
\label{Hlin}
\end{equation}
where $\hat a$ $(\hat a^\dagger)$ is the annihilation (creation) operator for the optical mode, and $\hat b$ $(\hat b^\dagger)$ for the mechanical mode. $\Delta$ is the detuning
between the input laser and the cavity, $\omega_{\textrm M}$ is the natural frequency of the mechanical oscillator, and $g$ is the light-enhanced coupling strength. Constant $\Delta$, $\omega_{\textrm M}$, and $g$ correspond to standard optomechanics.
Dissipation is characterised by a single optical damping rate $\kappa$, and an intrinsic mechanical damping rate, $\Gamma_{\mathrm M}$.
 In the present work we consider the effects of modulating
parameters such as $\Delta$, $\omega_{\textrm M}$ and $g$:
\begin{equation}
{\hat{H}(t)} = -\Delta(t){\hat a}^\dagger {\hat a} +
            \omega_{\textrm M}(t) {\hat b}^\dagger {\hat b} +
            g(t) ({\hat a}^\dagger+ {\hat a}) ({\hat b}^\dagger+ {\hat b}).
\label{Hlin2}
\end{equation}

\begin{figure}[ht!]
\begin{center}
{\includegraphics[width=3.3in]{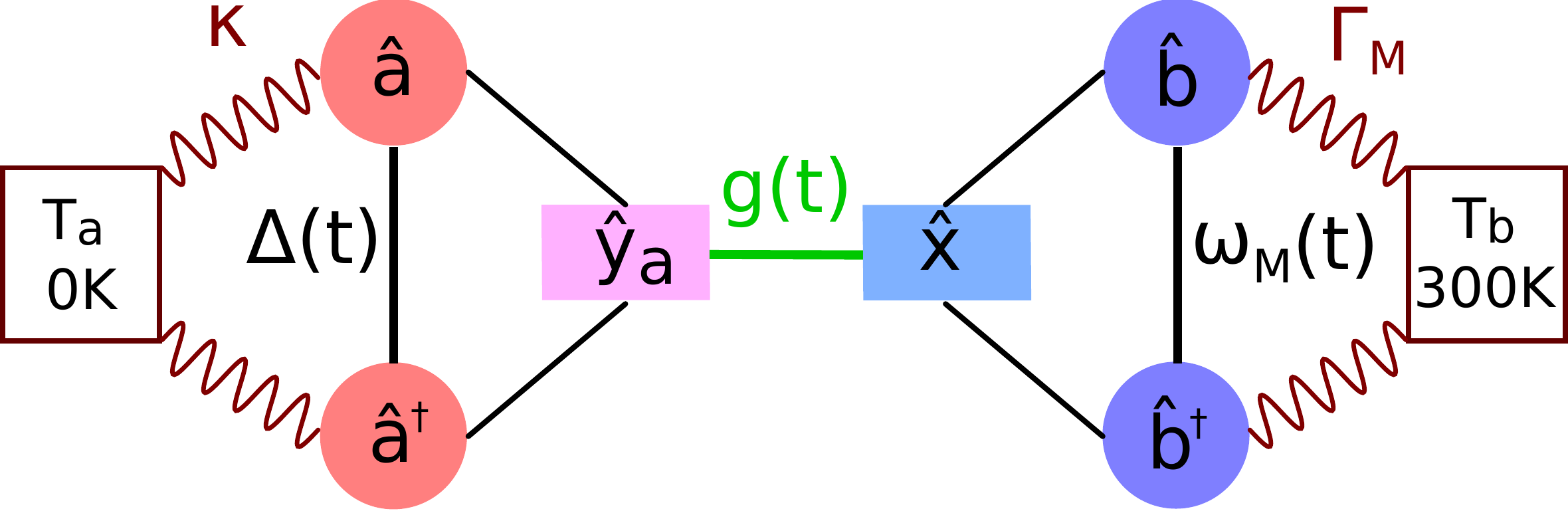}}
\end{center}
\caption{Schematic of the modulated optomechanical Hamiltonian.  $\hat a, \hat a^\dagger$ and $\hat b, \hat b^\dagger$ represent the optical and mechanical modes. The optical mode is driven by a strong coherent field, resulting in a linearized optomechanical coupling $g$ that couples the optical amplitude quadrature $\hat y_{\textrm a}(t) = \frac{1}{\sqrt 2} [\hat a(t) + \hat a^\dagger(t)]$ with the position quadrature $\hat x(t) = \frac{1}{\sqrt 2}[\hat b(t) + \hat b^\dagger(t)]$. The system operators are coupled to their respective baths by $\kappa$ for the zero-temperature optical bath, and by $\Gamma_{\textrm M}$ for the mechanical bath at $300 K$ leading to damping and dissipation. While for standard optomechanics
$g, \omega_{\textrm M}, \Delta$ are constant, we investigate here solutions for set-ups where they are harmonically modulated.}
\label{Fig1}
\end{figure}

Several previous theoretical studies of modulated optomechanics were motivated by the quest to overcome the Standard Quantum Limit (SQL) by measuring a single quadrature of the mechanical oscillator \cite{Caves1980}.
To date, two different ways to do single-quadrature detection have been proposed: one considered modulations of the optomechanical coupling $g$ to perform back-action evasion (BAE) measurements \cite{Caves1980, Clerk2008}, while the other considered modulation of $\omega_{\textrm M}$ to perform detuned mechanical parametric amplification (DMPA) \cite{Szorkovsky2014.Bow,Levitan2016.Clk}.  Closely related schemes to generate mechanical squeezing are also of much interest. Modulation of the cavity field at $2 \omega_{\mathrm M}$ results in amplification of one quadrature and to squeezing of the conjugate quadrature \cite{Mari2009.Eist, Kronwald2013}. The above studies all considered modulation either at or close to $\omega_{\textrm M}$ (resonant or near-resonant); or modulation at a multiple (usually twice) of $\omega_{\textrm M}$ \cite{Mari2009.Eist, Malz2016.Nunn}. In addition, they considered modulation of { \em either} the optomechanical coupling $g(t)$  \cite{Mari2009.Eist,Malz2016.Nunn} or of the spring constant \cite{Szorkovsky2014.Bow,Levitan2016.Clk}.

In this paper we revisit periodically modulated optomechanics by analysing
a system studied in \cite{Aranas2016.Mont}  which involves not only simultaneous modulation of {\em  both} $g$ and   $\omega_{\textrm M}$,
but  also {\em far off-resonant} modulation at frequencies $\ll\omega_{\textrm M}$.  That study was motivated by the need to understand the distinctive
optical sideband structure of the measured spectra from levitated nanoparticles in hybrid optical-electric traps \cite{Millen2015,Fonseca2016.Bark}.
In \cite{Aranas2016.Mont} an approximate but analytical solution was obtained for the quantum noise spectra of the optical field and mechanical displacement. The method produced closed form expressions which successfully reproduced experimental spectral features, but only for the case of weak modulations.

Quantum noise spectra for comparison with experiments or theory are obtained by transforming the corresponding quantum Langevin equations into frequency space; in the standard optomechanical case, these are frequently solved only for the individual mode of interest. However, a rather useful technique arises from so-called linear amplifier models \cite{Clerknoise,Botter2010.Kurn} which cast the equations in a matrix form: $\mathbf c(\omega) = \mathbf T(\omega) \mathbf c_{\textrm {in}}(\omega)$,
relating the input optical and mechanical noises to the field mode operator
outputs by means of a transfer matrix $\mathbf T$.
 Here the vector $\mathbf c(\omega) = \begin{pmatrix} \hat a(\omega) & \hat a^\dagger(\omega) & \hat b(\omega) & \hat b^\dagger(\omega) \end{pmatrix}^\mathsf T$
and the input vector $\mathbf c_{\rm {in}}(\omega)$ represent Gaussian noise baths.  We note that the  $\hat a(\omega)$ solutions here denote the intra-cavity field (the actual detected cavity output field  is then straightforwardly obtained using the input-output relation $\hat a_{\mathrm {out}}(\omega)=\hat a_{\mathrm {in}}(\omega)- \sqrt{\kappa} \hat a(\omega))$ \cite{Gardiner1985}.

Such matrix methods have been employed in previous studies of modulated optomechanics \cite{Mari2009.Eist,Malz2016.Nunn}. However, unlike the standard case, $\mathbf T$ couples frequencies which differ by multiples of the modulation frequency. Its dimension is infinite so truncation becomes necessary. In the present paper we identify two clear variants of the approach: in (i) the periodic Hamiltonian is expanded into Fourier series, and a covariance matrix equation is obtained in terms of frequency-shifted system operators. (ii) In \cite{Malz2016.Nunn}, a Floquet ansatz is used so steady-state solutions are assumed to be periodic. This results in a Langevin equation for each Fourier component of the system operators which can be arranged into a matrix equation in Fourier space. Although in \cite{Mari2009.Eist} a method equivalent to (i) was noted briefly, it has not previously been used to calculate quantum noise spectra.

We test the validity of the expressions of  \cite{Aranas2016.Mont} -- which we label method (iii) -- in thermal and quantum regimes. It is straightforward to show that the analytical expressions obtained by iterative solution in \cite{Aranas2016.Mont} are an approximate solution of method (i) for regimes where we may truncate the matrix $\mathbf T$ to the lowest few orders. We also investigate the subtle, but interesting, differences between the two Floquet/Fourier methods (i) and (ii) under certain assumptions in the input noise and detection methods. We prove that although the matrix equations are apparently different, the methods, in fact, yield equivalent power spectra.

In Sec. \ref{results}, we apply the formalism to the slowly modulated system in
\cite{Aranas2016.Mont} where the frequencies $g(t)$ and $\omega_{\textrm M}(t)$ are modulated at a frequency $\omega_{\textrm d} \ll \omega_{\textrm M}$. A destructive interference process that leads to complete cancellation of one of the displacement sidebands offers a very stringent test of the calculations. We verify the results for the intracavity spectra in the thermal regime by numerical simulation of the slowly modulated, semiclassical Langevin equations using a stochastic differential equation solver \cite{xmds}. We also calculate the quantum homodyne spectra in ponderomotive squeezing regimes (i.e. optical squeezing, not mechanical squeezing) even in the presence of strong modulations.

In Sec. \ref{discussion} we establish connections between the two methods (i) and (ii) to give a fuller picture of modulated optomechanical systems and their implications, in particular, by considering heterodyne detection of non-stationary spectral components which are usually inaccessible experimentally. Finally, we summarize and conclude in Sec. \ref{conclusion}.

\section{\label{theory} Theory: matrix methods for quantum noise spectra}

For compactness and generality, we can extend Eq. (\ref{Hlin}) into an $n$-mode quadratic Hamiltonian $\hat H(t) = \frac{1}{2} \mathbf c^{\sf T}(t) \mathbf H(t) \mathbf c(t)$, where the Hamiltonian \em matrix \em $\mathbf H(t)$ contains the coupling frequencies between the modes and $\mathbf c(t) = \begin{pmatrix} \hat c_{1}(t) & \hat c^\dagger_{1}(t) & \cdots & \hat c_{n}(t) & \hat c^\dagger_{n}(t) \end{pmatrix}^{\sf T}$ is a vector of $2 n$ system operators. The resulting Heisenberg's equation of motion is \cite{Serafini}:
\begin{equation}
\dot {\mathbf c}(t) = -\textrm i \sigma \mathbf H(t) \mathbf c(t),
\label{heisenberg}
\end{equation}
where we set $\hbar = 1$, and for bosonic ladder operators the canonical commutation relation (CCR) is
\begin{equation}
\sigma = [\mathbf c, \mathbf c^\dagger] = \bigoplus_{l=1}^n \begin{pmatrix} 1 & 0 \\ 0 & -1 \end{pmatrix}.
\label{CCR}
\end{equation}

Each of the $i$th mode of $\mathbf c(t)$ is coupled to an infinite bath with rate $\gamma_i$ which is described by a quantum Langevin equation:
\begin{equation}
\dot {\mathbf c}(t) = -\textrm i \sigma \mathbf H(t) \mathbf c(t) - \frac{\gamma}{2} \mathbf c(t)
+ \mathbf c_{\mathrm {in}}(t),
\label{QLE}
\end{equation}
where $\gamma = \sf{diag} \begin{pmatrix} \gamma_1 & \gamma_1 & \cdots & \gamma_n & \gamma_n \end{pmatrix}$, and the \em scaled \em input noise operators $\mathbf{c_{\textrm {in}}}(t) \equiv \begin{pmatrix} \sqrt \gamma_1 \hat c_{\textrm {in}, 1}(t) & \sqrt \gamma_1 \hat c^\dagger_{\textrm {in}, 1}(t) & \cdots & \sqrt \gamma_n \hat c_{\textrm {in}, n}(t) & \sqrt \gamma_n \hat c^\dagger_{\textrm {in}, n}(t) \end{pmatrix}^{\sf T}$. They are Gaussian noises which we assume to be delta-correlated:
\begin{eqnarray}
\left < \hat c_{\textrm {in}, i}(t) [\hat c_{\textrm {in}, i'}(t')]^\dagger \right > &=& (\bar n_i + 1) \delta_{ii'} \delta(t-t') \nonumber \\
\left < [\hat c_{\textrm {in}, i}(t)]^\dagger \hat c_{\textrm {in}, i'}(t') \right > &=& \bar n_i \delta_{ii'} \delta(t-t'),
\label{delta}
\end{eqnarray}
where we denote the $2i$th element of $\mathbf c_{\textrm {in}}(t)$ by $\hat c_{\textrm {in}, i}$. The mode occupancy $\bar n_i$ is set by the bath temperature. We further define a matrix of noise correlations in time:
\begin{flalign}
&\left < \mathbf c_{\textrm {in}}(t) [\mathbf c_{\textrm {in}}(t')]^\dagger \right >
\equiv \mathbf N \delta (t-t')
\label{noise} \\
&= \mathsf {diag} \begin{pmatrix}
\gamma_1 (\bar n_1 + 1) & \gamma_1 \bar n_1  & \cdots & \gamma_n (\bar n_n + 1)  & \gamma_n \bar n_n
\end{pmatrix} \delta (t-t'). \nonumber
\end{flalign}

In the case of a time-independent Hamiltonian $\mathbf H(t) = \mathbf H$, Eq. (\ref{heisenberg}) is diagonal in Fourier space:
\begin{equation}
\mathbf c(\omega) = \mathbf T(\omega) \mathbf{c_{\textrm {in}}(\omega)},
\label{i/o-standard}
\end{equation}
where the transfer matrix $\mathbf T(\omega) = \left (-\textrm i \omega \mathbf I + \textrm i \sigma \mathbf H + \frac{\mathbf \gamma}{2} \right )^{-1}$, $\mathbf I$ is the identity matrix, and our convention for the Fourier transform is such that: $[\mathbf c(\omega)]^\dagger = \int^{+\infty}_{-\infty} \mathrm d\omega \mathrm e^{-\mathrm i \omega t} [\mathbf c(t)]^\dagger$. Equation (\ref{i/o-standard}) underlines the essence of the linear amplifier model of standard optomechanics \cite{Botter2010.Kurn}: by working in frequency space we obtain the output noises from the input noises via simple matrix inversion.

The explicit time-dependence of $\mathbf H(t)$ -- slowly-modulated or otherwise -- prevents a straightforward application of the Fourier transform to obtain an input-output relation similar to Eq. (\ref{i/o-standard}). Nonetheless, one can apply Fourier techniques to Eq. (\ref{heisenberg}) in two ways: (i) by Fourier-expanding the Hamiltonian matrix, or (ii) by expanding both the Hamiltonian matrix and the system operators \cite{Malz2016.Nunn}. One can then obtain a linear system either of frequency-shifted operators or of the Fourier components of the operators. In the following text we show the equivalence of methods (i) and (ii) by deriving the power spectrum under two assumptions: 1.) input noises are Gaussian and stationary and 2.) no explicit time-dependence is introduced in the signal during detection.

\subsection{\label{shifted-operators} Method (i): Matrix equation of shifted operators}
First we express the periodic Hamiltonian matrix as a Fourier series: $\mathbf H(t) = \sum_{k \in \mathbb Z} \textrm {H}_k \textrm e^{\textrm i k \omega_{\textrm d}t}$. Equation (\ref{heisenberg}) becomes
\begin{equation}
\dot {\mathbf c}(t) = \left( -\textrm i \sigma \sum_k \textrm H_k \textrm e^{\textrm i k \omega_{\textrm d}t} -\frac{\mathbf \gamma}{2} \right ) \mathbf c(t)
+ \mathbf{c_{\textrm {in}}}(t),
\label{shifted}
\end{equation}
which in frequency space becomes
\begin{equation}
\left [ -\textrm i \omega \mathbf I + \frac{\gamma}{2} \right ] \mathbf c(\omega) =
-\textrm i \sigma \sum_k \textrm H_k \mathbf c(\omega + k \omega_{\textrm d}) + \mathbf c_{\textrm{in}}(\omega).
\label{unshifted-freq}
\end{equation}

Because of the time-dependence of $\mathbf H(t)$ the vector $\mathbf c(\omega)$ depends on $\mathbf c(\omega + k \omega_{\textrm d})$, preventing us from expressing Eq. (\ref{unshifted-freq}) as an input-output equation similar to Eq. (\ref{i/o-standard}). Instead, we consider the shifted equations:
\begin{eqnarray}
\left [ -\textrm i (\omega + s \omega_{\textrm d}) \mathbf I + \frac{\gamma}{2} \right ] \mathbf c(\omega + s \omega_{\textrm d}) = \nonumber \\
-\textrm i \sigma \sum_k \textrm H_k \mathbf c(\omega + (k + s) \omega_{\textrm d}) + \mathbf c_{\textrm{in}}(\omega + s \omega_{\textrm d})
\label{shifted-freq}
\end{eqnarray}
for each $k, s$. An input-output relation of the form $\mathbf c = \mathbf T \mathbf c_{\textrm {in}}$ can then be obtained for the modulated system:
\begin{widetext}
\begin{equation}
\begin{pmatrix} \vdots \\ \mathbf c(\omega + 2 \omega_{\textrm d}) \\ \mathbf c(\omega + \omega_{\textrm d}) \\ \mathbf c(\omega) \\ \mathbf c(\omega - \omega_{\textrm d}) \\ \mathbf c(\omega - 2 \omega_{\textrm d}) \\ \vdots \end{pmatrix} =
\begin{pmatrix}
\ddots &  &  & \vdots &  &  & \\
& \textrm X(\omega + 2\omega_{\textrm d}) & \textrm A_{-1} & \textrm A_{-2} & \textrm A_{-3} & \textrm A_{-4} & \\
& \textrm A_{1} & \textrm X(\omega + \omega_{\textrm d}) & \textrm A_{-1} & \textrm A_{-2} & \textrm A_{-3} & \\
\cdots & \textrm A_{2} & \textrm A_{1} & \textrm X(\omega) & \textrm A_{-1} & \textrm A_{-2} & \cdots \\
& \textrm A_{3} & \textrm A_{2} & \textrm A_{1} & \textrm X(\omega - \omega_{\textrm d}) & \textrm A_{-1} & \\
& \textrm A_{4} & \textrm A_{3} & \textrm A_{2} & \textrm A_{1} & \textrm X(\omega - 2\omega_{\textrm d}) & \\
&  &  & \vdots &  &  & \ddots
\end{pmatrix}^{-1}
\begin{pmatrix} \vdots \\ \mathbf c_{\textrm {in}}(\omega + 2 \omega_{\textrm d}) \\ \mathbf c_{\textrm {in}}(\omega + \omega_{\textrm d}) \\ \mathbf c_{\textrm {in}}(\omega) \\ \mathbf c_{\textrm {in}}(\omega - \omega_{\textrm d}) \\ \mathbf c_{\textrm {in}}(\omega - 2 \omega_{\textrm d}) \\ \vdots \end{pmatrix}
\label{matrix-freq}
\end{equation}
\end{widetext}
where the $n \times n$ matrix elements are
\begin{eqnarray}
\textrm A_{s} = \textrm i \sigma \textrm H_s
\label{Am} \\
\textrm X(\omega + s \omega_{\textrm d}) = -\textrm i (\omega + s \omega_{\textrm d}) \mathbf I + \textrm i \sigma \textrm H_0 + \frac{\gamma}{2}.
\label{Xm}
\end{eqnarray}
We denote $s$th row, $l$th column element of the transfer matrix as $\mathbf T_{sl}(\omega)$, with the central block being $[T^{-1}]_{00}(\omega) = X(\omega)$.

Our departure point to solve the measured power spectrum analytically is:
\begin{equation}
S_{\mathbf c \mathbf c^\dagger}(\omega) \equiv \lim_{T \rightarrow \infty} \left < \mathbf c(\omega) [\mathbf c(\omega)]^\dagger \right >.
\label{S-measured}
\end{equation}
where we have generalized for now to the case of operators, and frequency-space variables are understood to be gated Fourier transforms: $\mathbf c(\omega) = \frac{1}{\sqrt T} \int_0^T \mathrm dt \mathrm e^{\mathrm i \omega_{\mathrm d}t} \mathbf c(t)$. We note that Ref. \cite{Malz2016.Nunn} offers a different way to calculate the measured spectrum but we come back to this point later in Sec. \ref{results} C.

From Eq. (\ref{matrix-freq}) we know $\mathbf c(\omega) = \sum_{l \in \mathbb Z} \mathbf T_{0l}(\omega) \mathbf c_{\textrm {in}}(\omega - l \omega_{\rm d})$. Substituting this in Eq. (\ref{S-measured}) we obtain:
\begin{eqnarray}
S_{\mathbf c \mathbf c^\dagger}(\omega) &=& \lim_{T \rightarrow \infty} \sum_{l, l'} \mathbf T_{0l}(\omega) \left < \mathbf c_{\textrm {in}}(\omega - l \omega_{\textrm d}) \right . \nonumber \\
&\times& \left . [\mathbf c_{\textrm {in}}(\omega - l' \omega_{\textrm d})]^\dagger \right > [\mathbf T_{l'0}]^*(\omega)
\label{S_shifted}
\end{eqnarray}
It follows from Eq. (\ref{noise}) (proof in Appendix \ref{appA}) that
\begin{equation}
\lim_{T \rightarrow \infty} \left < \mathbf c_{\textrm {in}}(\omega - l \omega_{\textrm d}) [\mathbf c_{\textrm {in}}(\omega - l' \omega_{\textrm d})]^\dagger \right > = \mathbf N \delta_{ll'}.
\label{kronecker}
\end{equation}
Therefore,
\begin{equation}
S_{\mathbf c \mathbf c^\dagger}(\omega) = \sum_{l \in \mathbb Z} \mathbf T_{0l}(\omega) \mathbf N [\mathbf T_{l0}]^*(\omega).
\label{Scc}
\end{equation}

In the solution above the Hamiltonian matrix is Fourier expanded while the system operators are left as is, leading to a matrix equation of shifted operators.

\subsection{\label{fourier-modes} Method (ii): Matrix equation of Fourier modes}
In an alternative derivation \cite{Malz2016.Nunn} we expand both the Hamiltonian matrix and the system operators in a Fourier series. Let $\mathbf H(t) = \sum_{k \in \mathbb Z} \textrm H_k \textrm e^{\textrm i k \omega_{\textrm d}t}$ and $\mathbf c(t) = \sum_{l \in \mathbb Z} \mathbf c^{(l)}(t) \textrm e^{\textrm i l \omega_{\textrm d}t}$. Then starting from Eq. (\ref{QLE}) we use the relation $\mathbf H(t) \mathbf c(t) = \sum_k H_k \mathbf c(t) \textrm e^{\textrm i k \omega_{\textrm d}t} = \sum_k H_k \sum_l \mathbf c^{(l - k)} \textrm e^{\textrm i l \omega_{\textrm d} t}$ to arrive at:
\begin{eqnarray}
\sum_l \left [ \dot {\mathbf c}^{(l)}(t) + \left ( \textrm i l \omega_{\textrm d} \mathbf I + \frac{\gamma}{2} \right ) \mathbf c^{(l)}(t) \right ] \textrm e^{\textrm i l \omega_{\textrm d}t} \nonumber \\
= - \textrm i \sigma \sum_{l,k} \left [ H_k \mathbf c^{(l - k)}(t)
+ \mathbf c_{\textrm{in}}(t) \delta_{l,0} \right ] \textrm e^{\textrm i l \omega_{\textrm d}t}.
\end{eqnarray}
We identify a quantum Langevin equation for each Fourier mode:
\begin{eqnarray}
\left [ -\textrm i (\omega - l \omega_{\textrm d}) \mathbf I + \frac{\gamma}{2} \right ]
\mathbf c^{(l)}(\omega) \nonumber \\
= -\textrm i \sigma \sum_k H_k \mathbf c^{(l - k)}(\omega)
+ \mathbf c_{\textrm{in}}(\omega) \delta_{l,0},
\label{Langevin-mode}
\end{eqnarray}
Here we have assumed stationary input noise and placed it into the zeroth Fourier component. In general, periodic input noises can be treated as well \cite{Malz2016.Nunn2}.
The coupled quantum Langevin equations can be written as an infinite-dimensional matrix equation:
\begin{widetext}
\begin{equation}
\begin{pmatrix} \vdots \\ \mathbf c^{(-2)}(\omega) \\ \mathbf c^{(-1)}(\omega) \\ \mathbf c^{(0)}(\omega) \\ \mathbf c^{(1)}(\omega) \\ \mathbf c^{(2)}(\omega) \\ \vdots \end{pmatrix} =
\begin{pmatrix}
\ddots &  &  & \vdots &  &  & \\
& \textrm X(\omega + 2 \omega_{\textrm d}) & \textrm A_{-1} & \textrm A_{-2} & \textrm A_{-3} & \textrm A_{-4} & \\
& \textrm A_{1} & \textrm X(\omega + \omega_{\textrm d}) & \textrm A_{-1} & \textrm A_{-2} & \textrm A_{-3} & \\
\cdots & \textrm A_{2} & \textrm A_{1} & \textrm X(\omega) & \textrm A_{-1} & \textrm A_{-2} & \cdots \\
& \textrm A_{3} & \textrm A_{2} & \textrm A_{1} & \textrm X(\omega - \omega_{\textrm d}) & \textrm A_{1} & \\
& \textrm A_{4} & \textrm A_{3} & \textrm A_{2} & \textrm A_{1} & \textrm X(\omega - 2 \omega_{\textrm d}) & \\
&  &  & \vdots &  &  & \ddots
\end{pmatrix}^{-1}
\begin{pmatrix} \vdots \\ 0 \\ 0 \\ \mathbf c_{\textrm {in}}(\omega) \\ 0 \\ 0 \\ \vdots \end{pmatrix},
\label{matrix-mode}
\end{equation}
\end{widetext}
with the same transfer matrix as in Eq.(\ref{matrix-freq}).

From Eq. (\ref{matrix-mode}), the $l$th Fourier mode $\mathbf c^{(l)}(\omega) = \mathbf T_{l0}(\omega) \mathbf c_{\textrm{in}}(\omega)$, so $\mathbf c(\omega) = \sum_{l \in \mathbb Z} \mathbf c^{(l)}(\omega + l\omega_{\textrm d})$. Using Eq. (\ref{S-measured}), we then proceed to calculate the measured power spectrum:
\begin{eqnarray}
S_{\mathbf c \mathbf c^\dagger}(\omega) &=&  \lim_{T \rightarrow \infty} \sum_{l, l'} \mathbf T_{l0}(\omega + l \omega_{\mathrm d}) \left < \mathbf c_{\textrm {in}}(\omega + l \omega_{\mathrm d}) \right . \nonumber \\
&\times& \left . [\mathbf c_{\textrm {in}}(\omega + l' \omega_{\mathrm d})]^\dagger \right > [\mathbf T_{0l'}(\omega + l' \omega_{\mathrm d})]^*
\end{eqnarray}
Using the Kronecker delta correlation in Eq. (\ref{kronecker}), we obtain:
\begin{equation}
S_{\mathbf c \mathbf c^\dagger}(\omega) = \sum_{l \in \mathbb Z} \mathbf T_{l0}(\omega + l \omega_{\mathrm d}) \mathbf N [\mathbf T_{0l}(\omega + l \omega_{\mathrm d})]^*.
\label{S-mode}
\end{equation}

The infinite matrix $\mathbf T$ and its inverse have diagonals that are invariant (up to a frequency displacement) with respect to an equal shift in the row and column indices:
\begin{equation}
\mathbf T_{ll}(\omega) = \mathbf T_{l+n, l+n}(\omega + n \omega_{\textrm d}).
\label{translation}
\end{equation}
This translation property of $\mathbf T$ is a crucial feature that we will invoke throughout the paper. Shifting the indices of Eq. (\ref{S-mode}) by $-l$ we see that Eqs. (\ref{Scc}) and (\ref{S-mode}) are equivalent. We then conclude that methods (i) and (ii) yield equivalent power spectra, a key result of this work.

\subsection {\label{iterative} Method (iii): Iterative analytical solution}

Method (iii) is a solution obtained in \cite{Aranas2016.Mont, Aranas2017.Mont} to
a system where the coupling strength and mechanical spring constant are simultaneously modulated.
This challenging scenario was motivated by the need to analyze and understand the underlying dynamics for a particular experimental set-up with levitated nanoparticles in an optical cavity.

Levitated optomechanics offers the prospect of full decoupling from environmental heating and decoherence using nanoparticles trapped only by optical fields. This necessitates operation at ultra-high vacuum
$\sim 10^{-8} \, \textrm{mbar}$. However, previous studies identified a particle loss mechanism as the pressure is lowered past 1 mbar, presenting a major technical bottleneck. One solution was to incorporate a Paul trap inside the optical cavity \cite{Fonseca2016.Bark} to create a hybrid electro-optical trap.

In addition to interesting nonlinear dynamics, the hybrid trap system exhibited characteristic split-sideband spectra. These were analyzed \cite{Aranas2016.Mont,Aranas2017.Mont} by considering a simultaneous and  out-of-phase excursion in $g(t) = 2 \bar g \sin{\omega_{\textrm d} t}$ and $\omega_{\textrm M}(t) = \bar \omega_{\textrm M} + 2 \omega_2 \cos{2 \omega_{\textrm d} t}$.

Further details of the method in \cite{Aranas2016.Mont, Aranas2017.Mont} are given in Appendix \ref{appB2}. However in brief, it is useful to compare the frequency solution of the Langevin equation for the optical field amplitude
$\hat y(t) = \frac{1}{\sqrt 2} \left [\hat a(t) + \hat a^\dagger(t) \right ]$ for the standard unmodulated optomechanical case:
\begin{eqnarray}
\hat y(\omega) = \textrm i  g \eta(\omega)  \hat x(\omega)
+ \sqrt{\kappa} \hat Y_{\textrm{in}}(\omega),
\label{stand}
\end{eqnarray}
with the case where the optomechanical coupling strength is modulated
as $g(t) = 2 \bar g \sin{\omega_{\textrm d} t}$:
\begin{eqnarray}
\hat y(\omega) = \textrm i \bar g \eta(\omega) \left [ \hat x(\omega + \omega_{\textrm d}) - \hat x(\omega - \omega_{\textrm d}) \right ]
+ \sqrt{\kappa} \hat Y_{\textrm{in}}(\omega)
\label{iterate}
\end{eqnarray}
where the $Y_{\textrm{in}}(\omega)$ represent cavity-filtered incident shot noise (see Appendix \ref{appB2}) and $\hat x(\omega)$ is the displacement of the mechanical oscillator.

We see that the only apparent significant change is to the mechanical displacement operators which are frequency shifted by the modulation. Hence it might be tempting to substitute standard optomechanics noise expressions for $\hat x(\omega)$ by simply shifting  $ \omega \to \omega \pm \omega_{\textrm d}$ and to directly solve the equation.

However, the most physically interesting effects \cite{Aranas2016.Mont} arise from cross-correlations
$\langle  \hat x(\omega + \omega_{\textrm d}) \hat x(\omega - \omega_{\textrm d})\rangle$
between the $\omega \pm \omega_{\textrm d}$ components, generated by the second ($2\omega_{\textrm d}$) modulation of $\omega_{\textrm M}(t)$.

An iterative analytical solution was developed for the operator $\hat x(\omega + \omega_{\textrm d}) - \hat x(\omega - \omega_{\textrm d})$ (see Appendix \ref{appB2}) which successfully reproduced experimental features, but remained accurate only for weak $g$ and $\omega_2$. It is straightforward to see by inspection that the expressions used for the iterative solution are the central rows of Eq. \ref{matrix-freq} for $c(\omega)$: in other words, the iterative method is simply an approximation to the shifted operator method.

\section{\label{results} RESULTS: Simulation of split-sideband spectra}

In this section we test and verify the expressions for the methods (i)-(iii) -- both the iterative and the full matrix solution -- against each other and against a numerical solution of the stochastic Langevin equations. Methods (i) and (ii) yield indistinguishable results. Both solutions show the same convergence properties in that they need to be truncated at a higher order as the modulations become stronger. To ensure invertibility and convergence, we truncate the matrix in Eq. (\ref{matrix-freq}) at an arbitrarily high odd dimension ($17 \times 17$ block matrices).

For the numerics we explicitly solved a set of stochastic Langevin equations
corresponding to the  semiclassical dynamics of the system where we replace each operator in Eq. (\ref{matrix-standard}) with its (in general complex) expectation value and its adjoint with the
corresponding  complex conjugates. The stochastic noises $c_{\mathrm {in}}$ have a Gaussian distribution with an average variance equal to the step size in the temporal propagation, such that $\left < c_{\mathrm {in}, i}(t) c^*_{\mathrm {in}, i'}(t') \right > = \left < c^*_{\mathrm {in}, i}(t) c_{\mathrm {in}, i'}(t') \right > = 2 \pi (\bar n + 1/2) \delta_{i, i'} \delta (t - t')$.

\subsection{Split-sideband spectra in the strong modulation regime}
\begin{figure}[h]
\begin{center}
{\includegraphics[width=3.5in]{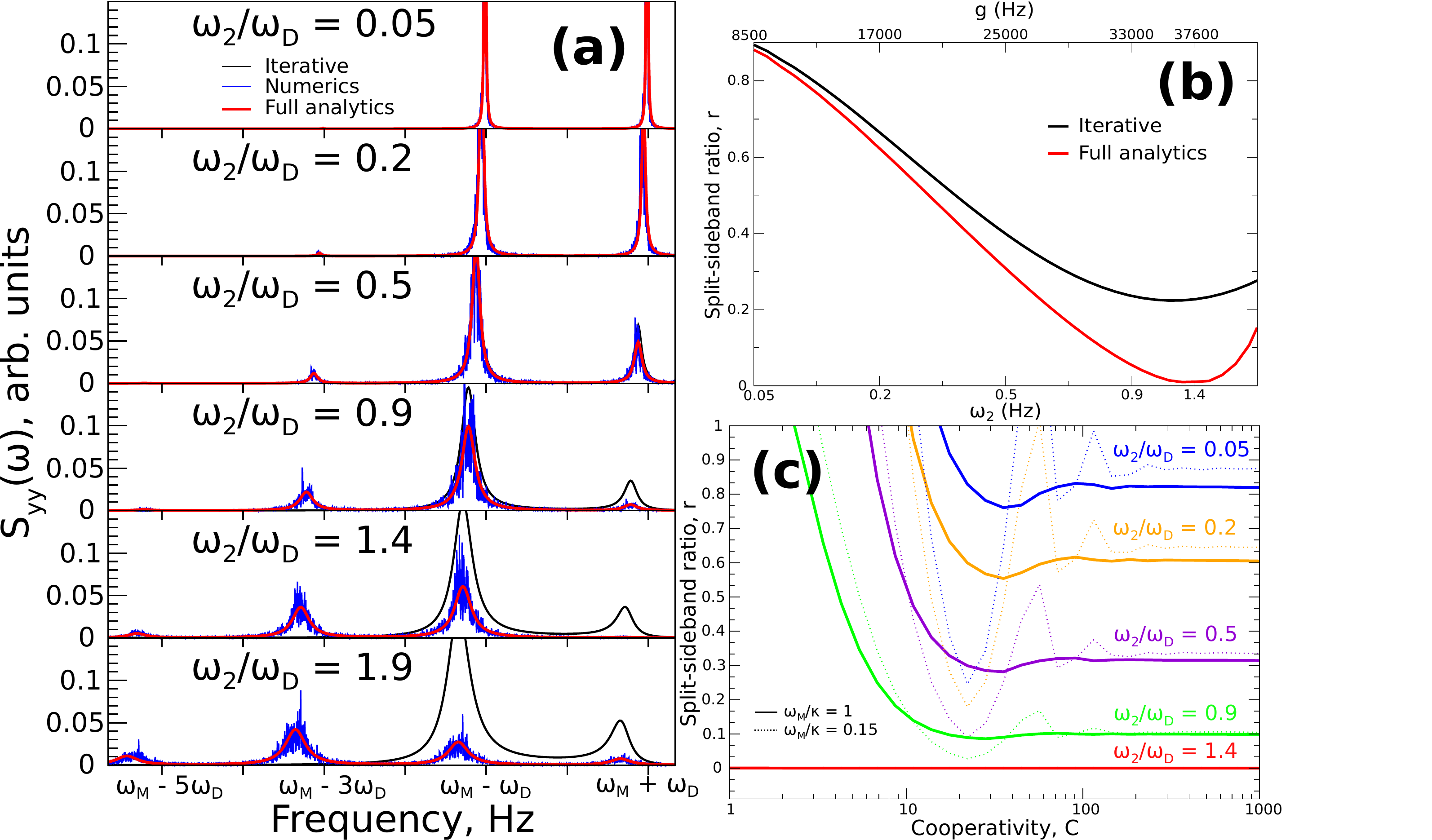}}
\end{center}
\caption{{\bf a.)} Comparison of the full analytical solution (red) with the iterative solution (black) of the cavity spectrum $S_{\textrm y \textrm y}(\omega)$ for different values of $\omega_2/\omega_{\textrm d}$, with $\omega_{\textrm d}$ fixed. The stochastic numerics (blue) are obtained by solving the first-order coupled Langevin equations using XMDS2. There is good agreement among the three, where we see one of the twin peaks is progressively suppressed, until around $\omega_2/\omega_{\textrm d} = 0.9$, where from this point onward the iterative solution fail to show further suppression. The full analytical spectra, on the other hand, agree very well with numerics - even showing higher-order sidebands. Parameters are: $\omega_{\textrm d}/\omega_{\textrm M} = 0.05$, $\Delta_2 = 0$, $\bar n_{\textrm b} = \frac{k}{\hbar \omega_{\textrm M}} 300 K$. {\bf b.)} Sideband ratio vs. $g$ and $\omega_{\mathrm M}$ for the same system in {\bf (a)}. Note the full analytical solution achieves the suppression point, after which the ratio bounces back to $R > 0$ as $\omega_2/\omega_{\textrm d}$ is further increased. The $g$ in {\bf (a)} changes with each $\omega_2/\omega_{\mathrm d}$ and is given here in the alternative axis. {\bf c.} Split-sideband ratio vs. cooperativity $C = \frac{4 g^2}{\kappa \Gamma_{\textrm M}}$ for $\omega_2/\omega_{\textrm d} =$ 0.05, 0.2, 0.5, 0.9, and 1.4; and for both sideband-resolved (solid, $\omega_{\textrm M}/\kappa = 1$) and otherwise (dotted, $\omega_{\textrm M}/\kappa = 0.15$). Parameters are: $\omega_{\textrm d}/\omega_{\textrm M} = 0.05$, $\Delta_2 = 0$, $\bar n_{\textrm b} = \frac{k}{\hbar \omega_{\textrm M}} 300 K$. Split-sideband resolution is ensured by the condition $\Gamma_{\textrm {opt}} \ll 2 \omega_{\textrm d} \leftrightarrow \frac{C \Gamma_{\textrm M}}{2 \omega_{\textrm d}} \ll 1$.}
\label{Fig2}
\end{figure}

Figure \ref{Fig2} compares methods (i)/(ii) with method (iii) as well as with the numerical simulation of the intracavity spectrum of the doubly-modulated system exhibiting the characteristic split-sideband separated by $2 \omega_{\textrm d}$ about $\omega_{\textrm M}$. To compare with previous studies \cite{Aranas2016.Mont}, each spectrum is parameterized by both $g$ and $\omega_2$. As was previously observed \cite{Aranas2016.Mont}, the ratio of the split-sidebands change as the parameter $\omega_2/\omega_{\textrm d}$ increases. Up to $\omega_2/\omega_{\textrm d} = 0.9$, all the three spectra exhibit progressively suppressed $\omega_{\textrm M} + \omega_{\textrm d}$ peak, and all show good agreement. From this point onward, however, the iterative solution fails to change the split-sideband ratio, while the full solution matches very well with the numerics, even going past the complete suppression point at $\omega_2/\omega_{\textrm d} \approx \sqrt 2$. We can also see this behaviour in Fig. \ref{Fig2}b where we plot the ratio of the split-sidebands as $g$ and $\omega_2$ increases. We also verify in Fig. \ref{Fig2}c that the split-sideband ratio persists regardless of the cooperativity and is only determined by $\omega_2/\omega_{\textrm d}$. Depending on $\kappa/\omega_{\textrm M}$, the split-sideband ratio may fluctuate before reaching a constant value. The higher the $\omega_2/\omega_{\textrm d}$ the lower cooperativity is required to reach a constant ratio, so at the suppression point $r \approx 0$ for all C. We ensure split sidebands are well-resolved by choosing $\Gamma_{\textrm {opt}} \ll 2 \omega_{\textrm d}$.

A new result of the comparison with the full Fourier methods (i/ii) is
to provide a more accurate value of the point at which the second sideband
is fully suppressed: here  we observe the suppression point at $\omega_2/\omega_{\textrm d} \approx \sqrt{2}$. An earlier analysis of the based on the approximate method (iii) $\omega_2/\omega_{\textrm d} \sim 2$ \cite{Aranas2016.Mont}; however, that analysis of the low order iterative solution neglected the modification to the susceptibilities due to higher-order backactions. The second sideband remains very weak across the entire $\omega_2/\omega_{\textrm d} \sim 1$ to $2$ range so the underlying physical explanation remains valid. Curiously, an even simpler model, using a Bessel expansion of the modulations in the interaction Hamiltonian \cite{Aranas2017.Mont} also predicted
the more accurate $\omega_2/\omega_{\textrm d} \approx \sqrt{2}$ result.

\subsection{\label{probe} Optical squeezing in homodyne spectra}

Measured spectra detect the cavity output spectrum $\hat a_{\textrm {out}}(\omega) = \hat a_{\textrm {in}} - \sqrt{\kappa} \hat a(\omega)$, presenting additional interesting effects arising from correlations between the incoming noise and the intracavity field due to quantum backaction. In particular such correlations give rise to
ponderomotive squeezing and power spectrum values below the shot-noise floor near $\omega \approx \omega_{\textrm M}$.

The measured homodyne spectrum detects a single optical quadrature:
\begin{equation}
i_{\textrm {hom}}(t)= e^{i\phi}\hat a_{\textrm {out}}(t)+ e^{-i\phi}a_{\textrm {out}}^\dagger(t)
\end{equation}
and hence the measured power spectrum
 $S_{\textrm {hom}}(\omega)= \langle |i_{\textrm {hom}}(\omega)|^2 \rangle $,
has four components
 $S_{\textrm {hom}}(\omega) = \left < \hat a_{\textrm {out}}(\omega) [\hat a_{\textrm {out}}(\omega)]^\dagger \right > + \left < [\hat a_{\textrm {out}}(\omega)]^\dagger \hat a_{\textrm {out}}(\omega) \right > + \left < \hat a_{\textrm {out}}(\omega) \hat a_{\textrm {out}}(\omega) \right > \textrm e^{2 \textrm i \phi} + \left < [\hat a_{\textrm {out}}(\omega)]^\dagger [\hat a_{\textrm {out}}(\omega)]^\dagger \right > \textrm e^{-2 \textrm i \phi}$, and $\phi$ is the local oscillator phase ($\phi = 0$ for amplitude, and $\phi = \pi/2$ for phase quadrature).

Another advantage of the linear amplifier matrix formalism is that outputs the full covariance matrix, facilitating calculation of the homodyne spectra
which are constructed from several separate components.
 Usually, a probe mode different from the control beam is used for detection. When probe coupling is weak and $\Delta_{\textrm p} = 0$ it does not alter system dynamics but otherwise the probe could significantly couple to the oscillator motion regardless of the quadrature being measured. The matrix methods are extendable to any number of modes so we can easily incorporate probe dynamics.

\begin{figure}[h]
\begin{center}
{\includegraphics[width=3.5in]{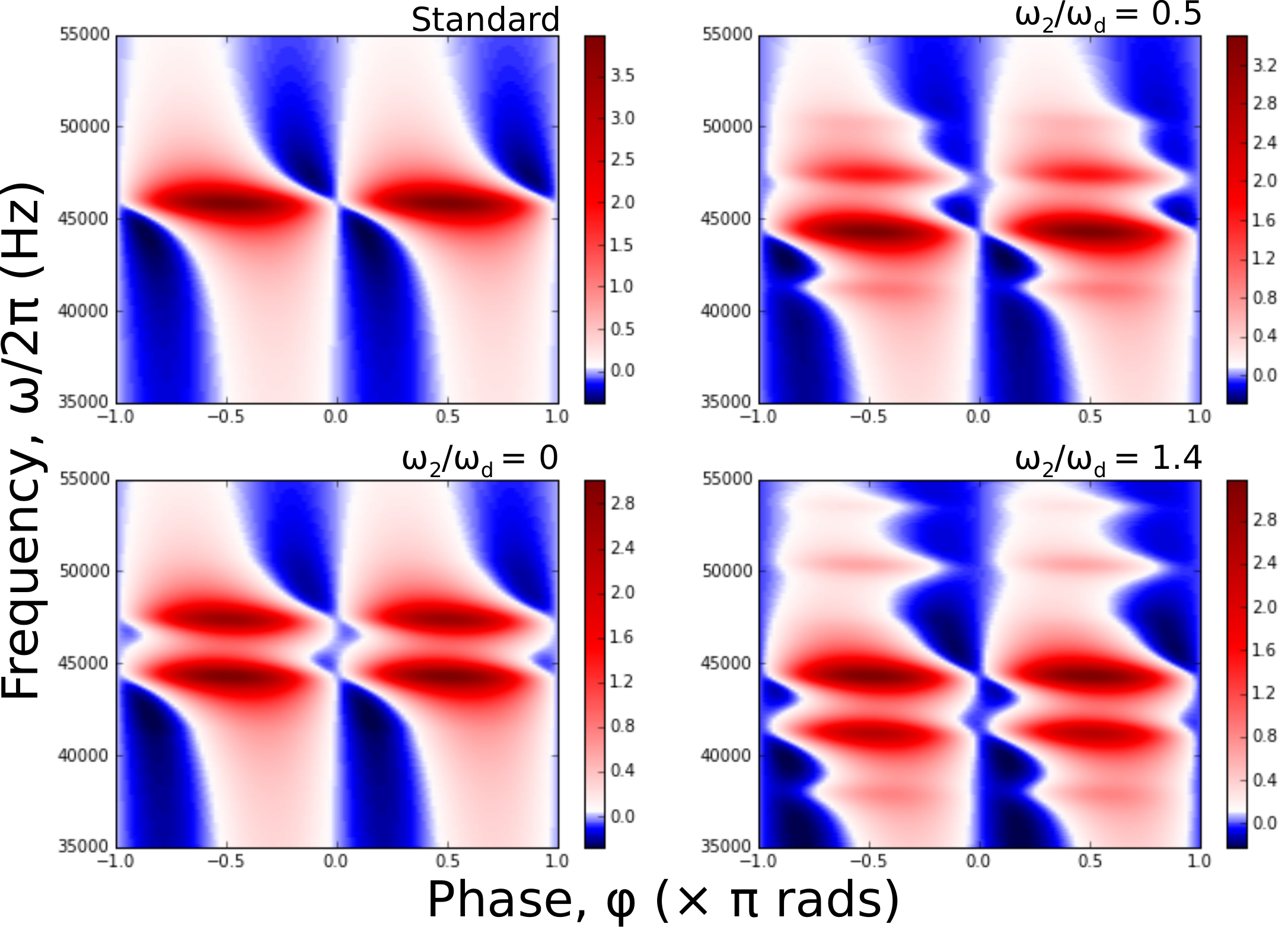}}
\end{center}
\caption{Color map of the homodyne spectra $S^\phi_{\textrm {hom}}(\omega)$ versus the local oscillator (LO) angle $\phi$ for the standard case, as well as the slowly-modulated cases with varying $\omega_2/\omega_{\textrm d}$. $g = 18.5$ kHz. We use two optical modes: the cooling mode at $\Delta = -\omega_{\textrm M}$ brings down the phonon occupation from $300$ K to $\bar n_{\textrm b} < 1$ while the probe mode at $\Delta_{\textrm p} = 0$ is used for readout. Both are at $\bar n_{\textrm a} = 0$ and $\Gamma_{\textrm M} = 2.3 \times 10^{-5}$. The blue (red) region indicate noise below (above) the imprecision floor. We get a flat spectrum for the amplitude quadrature ($\phi = 0$), while a twin-peak around $\omega_{\textrm M}$ for the phase quadrature $\phi = \pi/2$. We show the colormaps for the standard case, as well as for the slowly-modulated case for three different $\omega_2/\omega_{\textrm d}$. Not only do we see familiar regions of squeezing characteristic to standard optomechanics, but also squeezing between the twin-peaks. Maximum squeezing at $\approx 1$ dB (20\% below the noise floor) is achieved at $\phi = \pi/4$. At the suppression point $\omega_2/\omega_{\textrm d} \approx \sqrt 2$ regions of high backaction noise (red) are replaced by squeezing. The rest of parameters are the same as in Fig. \ref{Fig2}a.}
\label{Fig3}
\end{figure}

Figure \ref{Fig3} shows the color map of the quantum homodyne spectra for the standard case, as well as the modulated case for three different modulation strengths. Large regions of squeezing of up to $\approx 1$ dB (20 \% below the noise floor) can be observed for $0 < \phi < \pi/2$. The matrix method correctly replicates the squeezing profile of the standard case \cite{Purdy2013.Reg}. As expected for an on-resonance probe, the optical field shows no peaks at $\phi = 0$ while coupling most strongly with the mechanical oscillator at $\phi = \pi/2$. Optical squeezing at the mechanical frequency is impossible to see in standard optomechanics through homodyne detection, so sensing on-resonance will always be degraded by back-action noise, unless one performs a synodyne detection \cite{Buchmann2016.Kurn}, or introduce modulations within the system \cite{Clerk2008, Levitan2016.Clk}.

Adding a slow modulation in $g(t)$ allows the measurement of the cross-correlation $\langle  \hat x(\omega + \omega_{\textrm d}) \hat x(\omega - \omega_{\textrm d})\rangle$ that causes squeezing between the twin peaks. Introducing an additional periodicity in $\omega_{\mathrm M}(t)$ at $2 \omega_{\mathrm d}$ further increases the contribution of the cross-correlation. The result is a squeezed region that grows with $\omega_2/\omega_{\textrm d}$ until it completely suppresses backaction noise (red) on resonance for $\omega_2/\omega_{\textrm d} \approx \sqrt 2$. Such optical squeezing have been demonstrated for resonantly modulated optomechanical systems, but off-resonant modulated optomechanical systems could possibly offer a novel way of exploiting cross-correlations for quantum sensing.

\section{\label{discussion} Discussion}
Although we have shown that both methods (i) and (ii) give the same results for both intracavity and homodyne-detected power spectra, we now investigate whether the equivalence holds for more general types of spectra. In particular, we discuss heterodyne detection of modulated optomechanical systems.

\subsection{\label{validity} Connections between methods (i) and (ii)}

In summary, for both methods a set of output field modes is obtained from a set of input noises by the action of a transfer matrix $\mathbf T$. However, in method (i) the output field operator $\mathbf c(\omega)$ originates from multiple, frequency-shifted input noise components $ c_{\mathrm {in}}(\omega+l \omega_{\mathrm d})$. In  method (ii), in contrast, the dynamical operators were decomposed into a Fourier series $\mathbf c(t) = \sum_{l \in \mathbb Z} \mathbf c^{(l)}(t) \textrm e^{\textrm i l \omega_{\textrm d}t}$. In this case
these components $\mathbf c^{(l)}(\omega + l \omega_{\mathrm d})$ originates from the effect of the transfer matrix on a {\em single} input noise component $ c_{\mathrm {in}}(\omega)$.

To investigate these differences, we revisit once more the measured power spectra by re-writing Eq. (\ref{S-measured}) in terms of the autocorrelation function \cite{Bowenbook}:
\begin{flalign}
&\lim_{T \rightarrow \infty}\left < \mathbf c(\omega) [\mathbf c(\omega)]^\dagger \right > \nonumber \\
&\equiv \lim_{T \rightarrow \infty} \frac{1}{T}\int_0^T \mathrm dt \int_0^T d\tau \mathrm e^{\mathrm i \omega \tau} \left <\mathbf c(t+\tau) [\mathbf c(t)]^\dagger \right > \nonumber \\
&= \lim_{T \rightarrow \infty} \int_{-T/2}^{T/2} \mathrm dt S(\omega,t)
\label{SPSD}
\end{flalign}
where $S(\omega, t)$ is defined as the Fourier transform of the autocorrelation function. For an ordinary (unmodulated) optomechanical system, the stationarity (i.e. time-translation invariance) of the stochastic process leads to the Wiener-Khinchin theorem:
$\lim_{T \rightarrow \infty}\left < \mathbf c(\omega) [\mathbf c(\omega)]^\dagger \right >  = S(\omega,0)$.

In method (ii), the periodic modulation of the $\mathbf c(t)$ naturally implies the periodic
modulation of $S(\omega,t)$:
 \begin{equation}
 S(\omega, t) = \sum_{l \in \mathbb Z} S^{(m)}(\omega) \mathrm e^{\mathrm i m \omega_{\mathrm d}t}
\end{equation}
and in \cite{Malz2016.Nunn} it was shown that the measured spectrum is
the zeroth-order component $S^{(0)}(\omega)$.

Although the higher order spectral terms $S^{(m)}(\omega)$ appear to be experimentally inaccessible, we show below that these $|m| > 0 $ contributions may be measured using heterodyne detection with a beat frequency $2 \Omega = n \omega_{\mathrm d}$ resonant with the modulation. Hence the question arises as to how they can be calculated. It has been shown that $S^{(m)}(\omega)$ can be computed from the Fourier components of the operator using method (ii) \cite{Malz2016.Nunn}. In Appendix \ref{appD} we show that the higher spectral components are, in fact, straightforwardly related to cross-correlations between the method (i) operators:

\begin{equation}
\lim_{T \rightarrow \infty} \left < \mathbf c(\omega) [\mathbf c(\omega + m \omega_{\mathrm d})]^\dagger \right > = S^{(m)}(\omega)
\end{equation}
and hence higher order components of the spectrum are also obtainable from method (i).

All methods analyzed here assumed input noises which are stationary and delta-correlated, a standard assumption in optomechanics. Reference \cite{Malz2016.Nunn2} considered also the particular case where
the noise input to a cavity is itself the cavity-filtered noise from another cavity (with modulated dynamics). In this case one might consider noise inputs of the form:
\begin{equation}
\mu_{\mathrm {in}}(t) = \sum_{l \in \mathbb Z} \mathbf c^{(l)}_{\mathrm {in}}(t) \mathrm e^{\mathrm i l \omega_{\mathrm d} t},
\label{expand-noise}
\end{equation}
Even with periodic input noises, methods (i) and (ii) are still equivalent and we show this in Appendix \ref{appC}. Equation (\ref{expand-noise}) does not change the linearity of the equations in frequency space. The resulting frequency shifted noises $\mathbf c^{(l)}_{\mathrm {in}}(\omega + l \omega_{\mathrm d})$ can be accommodated by rearranging the matrix equation, though in this case one might find method (ii) more convenient.

\subsection {Measuring non-stationary spectrum components with heterodyne detection}

Heterodyne detection measures a rotating quadrature:
\begin{equation}
i_{\mathrm {het}}(t)= e^{i\phi+ \Omega t}\hat a_{\textrm {out}}(t)+ e^{-i(\phi+ \Omega t) }a_{\textrm {out}}^\dagger(t)
\end{equation}
and we take $\phi=0$ as the power spectrum is in general insensitive to $\phi$. Hence, in frequency space,
 $i_{\mathrm {het}}(\omega) = \hat a_{\mathrm {out}}(\omega + \Omega) + [\hat a_{\mathrm {out}}(\omega - \Omega)]^\dagger$.

In getting the power spectrum $S_{\mathrm {het}}(\omega) = \lim_{T \rightarrow \infty} \left < |i_{\mathrm {het}}(\omega)|^2 \right >$, intuition suggests that only the terms correlated at the same frequency will survive while the cross-correlations will vanish. Another way to look at this is through the time domain, where the heterodyne signal in time will give rise to a time-dependent autocorrelator, and the cross terms carrying $\pm \mathrm e^{2 \mathrm i \Omega t}$ will get averaged out in the Fourier transform \cite{Malz2016.Nunn}. Both viewpoints regarding the cancellation of cross correlations rely on the crucial fact that the noises are delta-correlated. However, on closer inspection, the cross correlations $\left < \hat a_{\mathrm {out}}(\omega - \Omega) \hat a_{\mathrm {out}}(\omega + \Omega)\right >$ and $\left < [\hat a_{\mathrm {out}}(\omega + \Omega)]^\dagger [\hat a_{\mathrm {out}}(\omega - \Omega)]^\dagger \right >$ can indeed be measured if the local oscillator frequency $\Omega$ is chosen appropriately. This is easy to show using method (i):
\begin{flalign}
&\lim_{T \rightarrow \infty} \left < \mathbf c(\omega + \Omega) [\mathbf c(\omega - \Omega)]^\dagger \right >
\nonumber \\
&= \lim_{T \rightarrow \infty} \sum_{l, l' \in \mathbf Z} \mathbf T_{0l}(\omega + \Omega)
\left < \mathbf c_{\mathrm {in}}(\omega + l \omega_{\mathrm d} + \Omega) \right . \nonumber \\
&\times \left . [\mathbf c_{\mathrm {in}}(\omega + l' \omega_{\mathrm d} - \Omega)]^\dagger \right >
[\mathbf T_{l'0}(\omega - \Omega)]^* \nonumber \\
&= \sum_{l \in \mathbf Z} \mathbf T_{0l}(\omega + \Omega) \mathbf N \mathbf [T_{l+n, 0}(\omega - \Omega)]^*
\label{cross}
\end{flalign}
The noise correlation in Eq. (\ref{noise}) forces $l' = l + n$, and also $n \equiv \frac{2 \Omega}{\omega_{\mathrm d}} \in \mathbb Z$. Such cross-correlations are useful in quantum sensing \cite{Kronwald2013, Buchmann2016.Kurn}, and Eq. (\ref{cross}) illuminates the interesting fact that, by introducing an appropriate phase reference $\Omega$ -- whether intrinsic to the system, or externally -- it becomes possible that a delta-correlated input noise (which vanishes if $\omega \neq \omega'$) can give rise to a non-zero correlation of output noises at different frequencies. In particular, we have shown how cross-correlations (and hence, how rotating parts of the cavity output spectrum) can be recovered naturally in modulated systems using heterodyne detection. The same idea has been applied on the level of rotating \em mechanical \em quadratures using the Fourier components of the periodic spectrum \cite{Malz2016.Nunn} which, we know from Appendix \ref{appD}, are equivalent to unequal-frequency cross correlations of shifted operators.

\section{\label{conclusion} Summary and conclusion}
We present three approaches to solving quantum noise spectra of periodically modulated optomechanical systems: we call these (i) shifted operators, (ii) Floquet, and (iii) iterative methods. We prove that methods (i) and (ii) yield equivalent spectra, while method (iii), is an analytical approximation to method (i).

We compare the equivalent methods (i)/(ii) with Langevin stochastic simulations of the doubly-modulated optomechanical Hamiltonian. The previously unexplored regime of slow but strong modulations in the optomechanical coupling and mechanical frequency provide a stringent test of the analytical methods. We demonstrate excellent agreement between methods (i)/(ii), confirming split-sideband suppression at $\omega_2/\omega_{\mathrm d} \approx \sqrt 2$. Method (iii), being effectively a low-order truncation of the transfer matrix of method (i), also shows good agreement up to a certain modulation amplitude.

We also predict \em resonant \em squeezing in the quantum regime for the doubly-modulated system as a result of enhanced cross-correlations in the shifted mechanical spectrum when $\omega_2/\omega_{\mathrm d} \approx \sqrt 2$. While squeezing at the mechanical frequency has been seen in other modulated schemes \cite{Buchmann2016.Kurn, Clerk2008}, we demonstrate possible new schemes for resonant squeezing in slowly-modulated set-ups.

Finally, we obtain a fuller picture of the periodic character of the spectra of the Langevin solutions by establishing an explicit connection between unequal-frequency correlations of shifted operators and the Fourier components of the periodic spectrum. We also show how cross-correlations (and hence, rotating components of the spectrum) are recovered by choosing the heterodyne local oscillator frequency to be resonant with the modulation of the optomechanical system.

\section{ACKNOWLEDGEMENTS} \label{acknowledgements}
We are grateful for insightful discussions with Andreas Nunnenkamp. EBA acknowledges fellowship support from Schlumberger Faculty For the Future program. MJA is thankful for a Stocklin-Selmoni studentship. DM acknowledges DTA studentship support by the UK Engineering and Physical Sciences Research Council (EPSRC) under Grant No. EP/M506485/1.

\appendix
\section{Frequency-space noise correlation in terms of Kronecker delta}\label{appA}
In this appendix we show that the noise correlation used to derive Eq. (\ref{Scc}) follows from the delta-correlation in time of Eq. (\ref{delta}).
\begin{flalign}
& \lim_{T \rightarrow \infty} \left < \mathbf c_{\mathrm {in}}(\omega + l \omega_{\mathrm d}) [\mathbf c_{\mathrm {in}}(\omega + l' \omega_{\mathrm d})]^\dagger \right > \nonumber \\
&= \lim_{T \rightarrow \infty}  \left < \frac{1}{\sqrt T} \int^{T}_{0} \mathrm dt \mathrm e^{\mathrm i (\omega + l \omega_{\mathrm d}) t} \mathbf c_{\mathrm {in}}(t) \right . \nonumber \\
&\times \left . \frac{1}{\sqrt T} \int^{T}_{0} \mathrm dt' \mathrm e^{-\mathrm i (\omega + l' \omega_{\mathrm d}) t} [\mathbf c_{\mathrm {in}}(t')]^\dagger \right > \nonumber \\
&= \lim_{T \rightarrow \infty}  \frac{1}{T} \int^{T}_{0} \mathrm dt \mathrm e^{\mathrm i (\omega + l \omega_{\mathrm d}) t}
\int^{T}_{0} \mathrm dt' e^{-\mathrm i (\omega + l' \omega_{\mathrm d}) t'}
\left < \mathbf c_{\mathrm {in}}(t) [\mathbf c_{\mathrm {in}}(t')]^\dagger \right > \nonumber \\
&= \lim_{T \rightarrow \infty}  \frac{1}{T} \int^{T}_{0} \mathrm dt \mathrm e^{\mathrm i (\omega + l \omega_{\mathrm d}) t}
\int^{T}_{0} \mathrm dt' e^{-\mathrm i (\omega + l' \omega_{\mathrm d}) t'}
\mathbf N \delta(t - t') \nonumber \\
&= \mathbf N \lim_{T \rightarrow \infty} \frac{1}{T} \int^{T}_{0} \mathrm dt \mathrm e^{\mathrm i (l-l') \omega_{\mathrm d} t}
\nonumber \\
&= \mathbf N \delta_{ll'}
\label{corr}
\end{flalign}
We can also generalize to the case of different frequencies that may arise from an external drive during detection. Assuming a frequency difference $\omega_{\textrm {diff}}$ and delta-correlated Fourier components of the noise, equation (\ref{corr}) becomes:
\begin{eqnarray}
\lim_{T \rightarrow \infty} \left < \mathbf c^{(m)}_{\mathrm {in}}(\omega + l \omega_{\mathrm d}) [\mathbf c^{(m')}_{\mathrm {in}}(\omega + \omega_{\textrm {diff}} + l' \omega_{\mathrm d})]^\dagger \right > \nonumber \\
= \mathbf N \delta_{(l-l')\omega_{\mathrm d}, \omega_{\textrm {diff}}} \delta_{m, m'}.
\label{corr2}
\end{eqnarray}
So in the case of Eq. (\ref{cross}), we take $\omega_{\textrm {diff}} = 2\Omega$. The Kronecker delta forces $\omega_{\textrm {diff}}$ to be an integer multiple of $\omega_{\mathrm d}$. For $\omega_{\textrm {diff}} = 0$, Eq. (\ref{corr2}) simplifies to Eq. (\ref{corr}).

\section{Analysis of the slowly-modulated system} \label{appB}
In this appendix we apply the general formalism in Sec. \ref{theory} to analyze in detail the slowly-modulated optomechanical system used to model levitated nanoparticles in a hybrid electro-optical trap.

\subsection{Time-periodic Langevin equations} \label{appB1}
Let $\mathbf c(t) \equiv \begin{pmatrix} \hat a(t) & \hat a^\dagger(t) & \hat b(t) & \hat b^\dagger(t) \end{pmatrix}^{\sf T}$ and denote the $2l$th element of $\mathbf c(t)$ by $\hat c_l$ so that $\hat c_1 \equiv \hat a$ and $\hat c_2 \equiv \hat b$. The optical and mechanical modes are coupled to their baths at $\kappa$ and $\Gamma_{\textrm M}$, respectively so $\gamma = \sf{diag} \begin{pmatrix} \kappa & \kappa & \Gamma_{\textrm M} & \Gamma_{\textrm M}\end{pmatrix}$. After symmetrising Eq. (\ref{Hlin}) and using the CCR Eq. (\ref{CCR}), Eq. (\ref{heisenberg}) for the optomechanical system is, explicitly,

\begin{widetext}
\begin{equation}
\begin{pmatrix} \dot{\hat a}(t) \\ \dot{\hat a}^\dagger(t) \\ \dot{\hat b}(t) \\ \dot{\hat b}^\dagger(t) \end{pmatrix} =
\begin{pmatrix}
\textrm i \Delta(t) - \frac{\kappa}{2} & 0 & \textrm i g(t) & \textrm i g(t) \\
0 & -\textrm i \Delta(t) - \frac{\kappa}{2} & -\textrm i g(t) & -\textrm i g(t) \\
\textrm i g(t) & \textrm i g(t) & -\textrm i \omega_{\textrm M}(t) - \frac{\Gamma_{\textrm M}}{2} & 0 \\
-\textrm i g(t) & -\textrm i g(t) & 0 & \textrm i \omega_{\textrm M}(t) - \frac{\Gamma_{\textrm M}}{2} \end{pmatrix}
\begin{pmatrix} \hat a(t) \\ \hat a^\dagger(t) \\ \hat b(t) \\ \hat b^\dagger(t) \end{pmatrix}
+ \begin{pmatrix} \sqrt \kappa \hat a_{\textrm {in}}(t) \\ \sqrt \kappa \hat a^\dagger_{\textrm {in}}(t) \\ \sqrt \Gamma_{\textrm M} \hat b_{\textrm {in}}(t) \\ \sqrt \Gamma_{\textrm M} \hat b^\dagger_{\textrm {in}}(t) \end{pmatrix}.
\label{matrix-standard}
\end{equation}
\end{widetext}

\subsection{Iterative analytical method} \label{appB2}
We review the iterative method to obtain a quantum solution that is valid in the low order, as previously introduced in \cite{Aranas2016.Mont}. From Eq. (\ref{matrix-standard}), the time-domain Langevin equations for the system operators are:
\begin{eqnarray}
\dot{\hat a}(t) &=& \left [ \textrm i \Delta(t) + \frac{\kappa}{2} \right ] \hat a(t)
+ \textrm i g(t) \left [ \hat b(t) + \hat b^\dagger(t) \right ] + \sqrt{\kappa} \hat a_{\textrm {in}}(t)
\nonumber \\
\dot{\hat b}(t) &=& -\left [ \textrm i \omega_{\textrm M}(t) + \frac{\Gamma_{\textrm M}}{2} \right ] \hat b(t)
+ \textrm i g(t) \left [ \hat a(t) + \hat a^\dagger(t) \right ] \nonumber \\
&+& \sqrt{\Gamma_{\textrm M}} \hat b_{\textrm {in}}(t)
\label{iterative-time}
\end{eqnarray}

Let us consider the specific case of a slowly-modulated optomechanical system where
\begin{eqnarray}
g(t) &=& 2 \bar g \sin{\omega_{\textrm d} t} \nonumber \\
\omega_{\textrm M}(t) &=& \bar \omega_{\textrm M} + 2 \omega_2 \cos{2\omega_{\textrm d} t} \nonumber \\
\Delta(t) &=& \bar \Delta.
\label{parameters}
\end{eqnarray}
Defining $\hat x(t) = \frac{1}{\sqrt 2} \left [\hat b(t) + \hat b^\dagger(t) \right ]$ and $\hat y(t) = \frac{1}{\sqrt 2} \left [\hat a(t) + \hat a^\dagger(t) \right ]$, we obtain from Eq. (\ref{iterative-time}) the position and optical amplitude quadratures in frequency space, respectively:
\begin{eqnarray}
\hat x(\omega) &=& \textrm i \bar g \mu(\omega) \left [ \hat y(\omega + \omega_{\textrm d}) - \hat y(\omega - \omega_{\textrm d}) \right ] + \sqrt{\Gamma_{\textrm M}} \hat X_{\textrm{th}}(\omega) \nonumber \\
&+& \textrm i \omega_2 \mathcal G(\omega) \nonumber \\
\hat y(\omega) &=& \textrm i \bar g \eta(\omega) \left [ \hat x(\omega + \omega_{\textrm d}) - \hat x(\omega - \omega_{\textrm d}) \right ]
+ \sqrt{\kappa} \hat Y_{\textrm{in}}(\omega), \nonumber \\
\label{iterate_app}
\end{eqnarray}
where the optical and mechanical susceptibilities are
\begin{eqnarray}
\chi_{\textrm O}(\omega) &=& [-\textrm i(\omega + \bar \Delta) + \frac{\kappa}{2}]^{-1} \nonumber \\
\chi_{\textrm M}(\omega) &=& [-\textrm i(\omega - \bar \omega_{\textrm M}) + \frac{\Gamma_{\textrm M}}{2}]^{-1}
\nonumber \\
\mu(\omega) &=& \chi_{\textrm M}(\omega) - \chi^*_{\textrm M}(-\omega) \nonumber \\
\eta(\omega) &=& \chi_{\textrm O}(\omega) - \chi^*_{\textrm O}(-\omega).
\end{eqnarray}
The input noises are
\begin{eqnarray}
\hat X_{\textrm{th}}(\omega) &=& \chi_{\textrm M}(\omega) \hat b_{\textrm {in}}(\omega) + \chi^*_{\textrm M}(-\omega) \hat b^\dagger_{\textrm {in}}(\omega) \nonumber \\
\hat Y_{\textrm{in}}(\omega) &=& \chi_{\textrm O}(\omega) \hat a_{\textrm {in}}(\omega) + \chi^*_{\textrm O}(-\omega) \hat a^\dagger_{\textrm {in}}(\omega),
\end{eqnarray}
and the correction due to $\omega_{\textrm M}$ excursion is $\mathcal G(\omega) \equiv \chi_{\textrm M}(\omega + 2 \omega_{\textrm d}) \hat b(\omega + 2 \omega_{\textrm d}) + \chi_{\textrm M}(\omega - \omega_{\textrm d}) \hat b(\omega - 2 \omega_{\textrm d}) - h.c.$

To calculate the power spectral density (PSD) we need to express the system operators solely in terms of input noises. Note, however, from Eq.(\ref{iterate_app}) that the output vectors in $\omega$ not only depend on input noises at $\omega$ but also at system operators at $\omega \pm \omega_{\textrm d}$ and $\omega \pm 2 \omega_{\textrm d}$. Hence we shift the quantum Langevin equations:
\begin{eqnarray}
\hat x(\omega \pm n \omega_{\textrm d}) &=&
\pm \textrm i \bar g \left [ \hat y(\omega \pm (n+1) \omega_{\textrm d}) - \hat y(\omega \pm (n-1) \omega_{\textrm d}) \right ] \nonumber \\
&+& \sqrt{\Gamma_{\textrm M}} \hat X_{\textrm{th}}(\omega \pm n\omega_{\textrm d})
+ \textrm i \omega_2 \mathcal G(\omega \pm n\omega_{\textrm d}) \nonumber \\
\hat y(\omega \pm n \omega_{\textrm d}) &=&
\pm \textrm i \bar g \left [ \hat x(\omega \pm (n+1) \omega_{\textrm d}) - \hat x(\omega \pm (n-1) \omega_{\textrm d}) \right ] \nonumber \\
&+& \sqrt{\kappa} \hat Y_{\textrm{in}}(\omega \pm n \omega_{\textrm d}),
\label{iterate-shift}
\end{eqnarray}
and iteratively substitute in Eq. (\ref{iterate_app}) the shifted vectors $\hat x(\omega \pm n\omega_{\textrm d})$ and $\hat y(\omega \pm n\omega_{\textrm d})$, for any $n \in \mathbb{Z}$, as they arise. Once we have $\hat y(\omega) = \sum_{l,n} \mathcal A_{c_l}(\omega + n \omega_{\textrm d}) \hat c_{\textrm {in}, l}(\omega + n \omega_{\textrm d}) + \mathcal A_{c^\dagger_l}(\omega + n \omega_{\textrm d}) \hat c^\dagger_{\textrm {in}, l}(\omega + n \omega_{\textrm d})$, the power spectrum is simply $S_{yy}(\omega) = \sum_l | \mathcal A_{c_l}(\omega + n \omega_{\textrm d}) |^2 \bar n_l + | \mathcal A_{c^\dagger_l}(\omega + n \omega_{\textrm d}) |^2 (\bar n_l + 1)$.

As noises from higher orders are considered, the iterative method becomes increasingly accurate but equally cumbersome if done by hand. In the following we apply the method in Eq. (\ref{shifted-operators}) to the slowly modulated optomechanical system with $n = 2$.

\subsection{Matrix equation for the slowly-modulated system} \label{appB3}
Equation (\ref{matrix-freq}) is a general equation that computes the system operators from the input noises for any $n$-mode modulated optomechanical system. To get the equation for a slowly-modulated system we set $\mathbf c(\omega + m \omega_{\textrm d}) \equiv \begin{pmatrix} \hat a(\omega + m \omega_{\textrm d}) & \hat a^\dagger(\omega + m \omega_{\textrm d}) & \hat b(\omega + m \omega_{\textrm d}) & \hat b^\dagger(\omega + m \omega_{\textrm d}) \end{pmatrix}^{\sf T}$ and $\mathbf c_{\textrm {in}}(\omega) \equiv \begin{pmatrix} \sqrt \kappa \hat a_{\textrm {in}}(\omega) & \sqrt \kappa \hat a^\dagger_{\textrm {in}}(\omega) & \sqrt \Gamma_{\textrm M} \hat b_{\textrm {in}}(\omega) & \sqrt \Gamma_{\textrm M} \hat b^\dagger_{\textrm {in}}(\omega) \end{pmatrix}^{\sf T}$. Moreover, the matrix elements are derived from the Hamiltonian Eq. (\ref{Hlin}) and the parameters in Eq. (\ref{parameters}), using Eq. (\ref{Am}) and Eq. (\ref{Xm}):
\begin{eqnarray} \textrm {X}_n = \sf diag
\left ( \chi_{\textrm O}(\omega + n\omega_{\textrm d}) \, \, \chi^*_{\textrm O}(-\omega - n\omega_{\textrm d})
\nonumber \right . \nonumber \\ \left .
\chi_{\textrm M}(\omega + n\omega_{\textrm d}) \, \, \chi^*_{\textrm M}(-\omega - n\omega_{\textrm d}) \right )
\label{A0}
\end{eqnarray}
\begin{eqnarray}
\textrm A_{\pm 1} = \pm \bar g
\begin{pmatrix}
  &  & 1 & 1 \\
  &  & -1 & -1 \\ \hline
1 & 1 & &  \\
-1 & -1 & &
\end{pmatrix}
\label{A1} \\
\textrm A_{\pm 2} = \textrm i
\begin{pmatrix}
-\Delta_2 & 0& & \\
0 & \Delta_2 & & \\ \hline
 & & \omega_2 & 0 \\
 & & 0 & -\omega_2,
\end{pmatrix}
\label{A2}
\end{eqnarray}
$\textrm A_{|n|>2} = 0$ because we do not consider here modulations greater than $2 \omega_{\textrm d}$. We substitute (\ref{A0}) to (\ref{A2}) in the matrix equation (\ref{matrix-freq}) and calculate the power spectrum using Eq. (\ref{Scc}).

\section{Equivalence of the analytical methods for periodic input noises} \label{appC}
To generalize our proof of the equivalence of analytical methods (i) and (ii), we relax the condition of stationary input noises and consider periodic input noises of the form Eq. (\ref{expand-noise}). Then the vector of operators is $\mathbf c(\omega) = \sum_{m \in \mathbb Z} \mathbf T_{0m}(\omega) \mu_{\mathrm {in}} (\omega - m \omega_{\mathrm d}) = \sum_{m,l \in \mathbb Z} \mathbf T_{0m}(\omega) \mathbf c^{(l)}_{\mathrm {in}}(\omega + [l - m] \omega_{\mathrm d})$, and the power spectrum using the shifted operators approach is:
\begin{eqnarray}
S^{(\mathrm i)}_{\mathbf c \mathbf c^\dagger}(\omega)
&=& \lim_{T \rightarrow \infty} \sum_{m, m', l, l'}
\mathbf T_{0m}(\omega) \left < \mathbf c^{(l)}_{\mathrm {in}}(\omega + [l - m] \omega_{\mathrm d})
\right . \nonumber \\
&\times& \left . [\mathbf c^{(l')}_{\mathrm {in}}(\omega + [l' - m'] \omega_{\mathrm d})]^\dagger \right >
[\mathbf T_{m'0}(\omega)]^* \nonumber \\
&=& \lim_{T \rightarrow \infty} \sum_{m, m' ,l, l'}
\mathbf T_{0,m-l}(\omega) \left < \mathbf c^{(l)}_{\mathrm {in}}(\omega - m \omega_{\mathrm d})
\right . \nonumber \\
&\times& \left . [\mathbf c^{(l')}_{\mathrm {in}}(\omega - m' \omega_{\mathrm d})]^\dagger \right >
[\mathbf T_{m'-l',0}(\omega)]^* \nonumber \\
&=& \sum_{m ,l} \mathbf T_{0,m-l}(\omega) \mathbf N [\mathbf T_{m-l,0}(\omega)]^*,
\end{eqnarray}
where we have started with Eq. (\ref{S-measured}), shifted the summation indices, and used a slight generalization of the correlation in Eq. (\ref{kronecker}) for different Fourier components of the noise.
The power spectrum via Floquet method can likewise be obtained:
\begin{eqnarray}
S^{(\mathrm {ii})}_{\mathbf c \mathbf c^\dagger}(\omega) &=& \lim_{T \rightarrow \infty}
\sum_{m,m',l,l'} \mathbf T_{ml}(\omega + m \omega_{\mathrm d}) \left < \mathbf c^{(l)}_{\mathrm {in}} (\omega + m \omega_{\mathrm d}) \nonumber \right .
\\ &\times& \left . [\mathbf c^{(l')}_{\mathrm {in}} (\omega + m' \omega_{\mathrm d})]^\dagger [\mathbf T_{l'm'}(\omega + m' \omega_{\mathrm d})]^* \right > \nonumber \\
&=& \sum_{m,l} \mathbf T_{ml}(\omega + m \omega_{\mathrm d}) \mathbf N [\mathbf T_{lm}(\omega + m \omega_{\mathrm d})]^*.
\end{eqnarray}
By invoking the translation property of $\mathbf T$ in Eq. (\ref{translation}), we therefore conclude that, after an appropriate shifting of the summation indices, methods (i) and (ii) yield equivalent spectra even in the more general case of periodic input noises.

\section{Components of the periodic spectrum in terms of shifted operators} \label{appD}
We show the components of the periodic spectrum can also be calculated using the shifted operators approach where they have a new interpretation as cross-correlations of operators shifted at different frequencies.

As mentioned Sec. \ref{results} C, the assumption of the Floquet formalism is a periodic spectrum $S(\omega, t) = \sum_{m \in \mathbb Z} S^{(m)}(\omega) \mathrm e^{\mathrm i m \omega_{\mathrm d}t}$ with Fourier components \cite{Malz2016.Nunn}:
\begin{flalign}
&S^{(m)}(\omega) = \sum_{l} \int_{-\infty}^{+\infty} \frac{\mathrm d \omega'}{2 \pi} \left < \mathbf c^{(l)}(\omega + l \omega_{\mathrm d}) [\mathbf c^{(l-m)}(\omega')]^\dagger \right > \nonumber \\
&= \sum_l \int_{-\infty}^{+\infty} \frac{\mathrm d \omega'}{2 \pi} \mathbf T_{l0}(\omega + l \omega_{\mathrm d})
\left < \mathbf c_{\mathrm {in}}(\omega) [\mathbf c_{\mathrm {in}}(\omega')]^\dagger \right > \nonumber \\
&\times [\mathbf T_{0, l-m}(\omega' + l \omega_{\mathrm d})]^* \nonumber \\
&= \sum_l \mathbf T_{l0}(\omega + l \omega_{\mathrm d}) \mathbf N \mathbf T^*_{0,l-m}(\omega + l \omega_{\mathrm d})
\end{flalign}
where we have used Eq. (\ref{matrix-mode}) and the noise correlation of Eq. (\ref{noise}) expressed in frequency space.

Consider the cross-correlation of shifted operators from method (i):
\begin{flalign}
&\lim_{T \rightarrow \infty} \left < \mathbf c(\omega) [\mathbf c(\omega + m \omega_{\mathrm d})]^\dagger \right > \nonumber \\
&= \sum_{l \in \mathbb Z} \mathbf T_{0l}(\omega) \left < \mathbf c_{\mathrm {in}}(\omega - l \omega_{\mathrm d}) [\mathbf c_{\mathrm {in}}(\omega - l \omega_{\mathrm d})]^\dagger \right > [\mathbf T_{l, -m}(\omega)]^*
\nonumber \\
&= \sum_l \mathbf T_{0l}(\omega) \mathbf N [\mathbf T_{l, -m}(\omega)]^* = S^{(m)}(\omega)
\end{flalign}
where in the last line we have invoked the translation property of $\mathbf T$. We then see that the Fourier components of the Fourier spectrum can be calculated using method (i).

\end{document}